\documentclass[aps,prb,twocolumn,showpacs]{revtex4}

\usepackage[T1]{fontenc}
\usepackage{amsmath,amsfonts,amssymb,amsthm,bbm,bm}
\usepackage{graphicx}
\usepackage{color}

\allowdisplaybreaks

\begin{document}

\title{Fisher information approach to non-equilibrium phase transitions in quantum XXZ spin chain with boundary noise}

\author{Ugo Marzolino$^{a,b}$, Toma\v z Prosen$^b$}

\affiliation{
${}^a$ Ru\dj er Bo\v{s}kovi\'c Institute, HR-10000 Zagreb, Croatia \\
${}^b$ Department of Physics, Faculty of Mathematics and Physics, University of Ljubljana, Jadranska 19, SI-1000 Ljubljana, Slovenia
}

\date{\today}

\begin{abstract}
We investigated quantum critical behaviours in the non-equilibrium steady state of a $XXZ$ spin chain with boundary Markovian noise using the Fisher information.
The latter represents the distance between two infinitesimally close states, and its superextensive size scaling witnesses a critical behaviour due to a phase transition,
since all the interaction terms are extensive.
Perturbatively in the noise strength, we found superextensive Fisher information at anisotropy $|\Delta|\leqslant1$ and irrational $\frac{\arccos\Delta}{\pi}$ irrespective of the order of two non-commuting limits,
i.e. the thermodynamic limit and the limit of sending $\frac{\arccos\Delta}{\pi}$ to an irrational number via a sequence of rational approximants.
From this result we argue the existence of a non-equilibrium quantum phase transition with a critical phase $|\Delta|\leqslant1$.
From the non-superextensivity of the Fisher information of reduced states, we infer that this non-equilibrium quantum phase transition does not have local order parameters but has non-local ones, at least at $|\Delta|=1$.
In the non-perturbative regime for the noise strength, we numerically computed the reduced Fisher information which lower bounds the full state Fisher information, and is superextensive only at $|\Delta|=1$.
Form the latter result, we derived local order parameters at $|\Delta|=1$ in the non-perturbative case.
The existence of critical behaviour witnessed by the Fisher information in the phase $|\Delta|<1$ is still an open problem.
The Fisher information also represents the best sensitivity for any estimation of the control parameter, in our case the anisotropy $\Delta$,
and its superextensivity implies enhanced estimation precision which is also highly robust in the presence of a critical phase.
\end{abstract}

\maketitle

\section{Introduction}

One of the paradigms for non-equilibrium statistical physics consists in the study of non-thermalising noisy dynamics \cite{HenkelHinrichsenLubeck2008,HenkelHinrichsenLubeck2010}:
non-equilibrium phase transitions are non-analytic changes of non-equilibrium steady states (NESS).
This kind of transitions has a much richer phenomenology than equilibrium phase transitions because NESSes lack a universal description in terms of thermodynamic potentials.
From a methodological point of view, this situation results in a large variety of universality classes without general tools for their characterisation \cite{Odor2004,Lubeck2004}. For instance, algebraically decaying correlation functions are not peculiar of critical phenomena \cite{Prosen2010}. Also the specral gap of the Liouvillian, an open system generalisation of the Hamiltonian gap, may vanish in the thermodynamic limit for all parameters, with critical points resulting only in a faster convergence \cite{Prosen2008,Znidaric2015}.

The broad interest on non-equilibrium phase transitions and on the search, pursued in our approach, for universal tools to characterise them is also manifest from their emergence in a large variety of settings, from complex systems, both physical \cite{Woodcock1985,Chrzan1992,Schweigert1998,Blythe2002,Whitelam2014,Zhang2015} and biological \cite{Egelhaaf1999,Marenduzzo2001,BarrettFreeman2008,Woo2011,Mak2016,Battle2016}, to social sciences and economics \cite{Recher2000,Llas2003,Baronchelli2007,Scheffer2012}. Furthermore, quantum-like models have been developed to fit phenomena in social sciences and economics \cite{Khrennikov2010}.

For quantum systems, dynamics with Markovian noise are represented by Lindblad master equations \cite{BreuerPetruccione2002,BenattiFloreanini2005}. Recently, many investigations enlightened complex and critical behaviours of quantum NESSes \cite{Diehl2008,Prosen2008,Diehl2010,DallaTorre2010,Prosen2010,Heyl2013,Ajisaka2014,
Znidaric2015,Vajna2015,Dagvadorj2015,Bartolo2016,Jin2016,Roy2017,Fink2017,Fitzpatrick2017}. An interesting paradigmatic master equation consisting of an anisotropic Heisenberg (XXZ) spin chain driven with an unequal noise at its boundaries has a non-trivial steady state 
with transitions manifested in transport properties \cite{Prosen2015}.

Exactly solvable models are forming one of the main pillars of classical statistical mechanics, both in and out of equilibrium. Among important general concepts which are amenable to exact solutions, are the 
non-equilibrium steady states (NESS), important nontrivial examples of which are the simple exclusion processes with boundary driving \cite{derrida}. Similar does not yet hold for quantum statistical mechanics, as the number of exact solutions for interacting models, in particular out of equilibrium, is very limited. The example of boundary driven XXZ model is one of the very few. Nevertheless, the behaviour of non-equilibrum partition function for a few other models that can be exactly solved using a similar boundary noise protocol (e.g., boundary driven Fermi-Hubbard model and an integrable $SU(3)$ chain \cite{Prosen2015}) is qualitatively identical to the one for the isotropic Heisenberg model. This leads us to believe that the boundary driven XXZ model discussed here may represent an important out-of-equilibrium universality class and the same type of phase transition may later be seen in other models. This may not be related to integrability, but in non-integrable systems the numerical analysis required to apply our approach in the NESS will be much harder.

Many equilibrium phase transitions are detected by the Bures metric, also known as fidelity susceptibility \cite{Gu2010}. The latter is proportional to the Fisher information \cite{Helstrom1976,Holevo1982,Paris2009,BengtssonZyczkowski2006} except in the presence of pathological behaviours however producing only removable singularities \cite{Safranek2016}. While this quantity reduces to standard susceptibilities for thermal phase transitions \cite{Janyszek1990,Ruppeiner1995,Quan2009,Marzolino2013,Marzolino2015}, it represents a more sophisticated tool for
quantum phase transitions (QPT), both symmetry-breaking \cite{CamposVenuti2007,Zanardi2007-2} and topological \cite{Yang2008} ones.
It is worth to mention another non-standard approach to phase transitions in equilibrium statistical mechanics and in chaotic dynamics, which is based on topological changes of isoenergetic manifolds in the phase space \cite{Pettini2007,Kastner2008}.
For non-equilibrium steady-state quantum phase transitions (NESS-QPT), as will be discussed here, the study of Fisher information is in the very early stage \cite{Banchi2014,Marzolino2014,Marzolino2016-2}.

The rationale of our approach relies upon the geometric interpretation of the Fisher information as the distance between two infinitesimally close states with respect to a varying control parameter. Indeed, when all interaction terms are extensive, superextensive metric implies instability with respect to small changes, e.g. due to critical points separating different phases.
In this paper, we exemplify this approach with deep characterisations of the NESS-QPT in the XXZ chain with boundary noise.
The above geometric interpretation provides a universal and unifying approach for both equilibrium and non-equilibrium, and possibly unknown, phase transitions, with a clear advantage over the aforementioned non-universal tools.

We also investigate, to best of our knowledge, original relations between non-local or local order parameters and the Fisher information of the full state or of reduced states, respectively. This relation is general and does not depend on the model and, as such, can be applied to any phase transition detected by the Fisher information. Moreover, it replies upon the Cram\'er-Rao bound, i.e. a result from estimation theory \cite{Helstrom1976,Holevo1982,Paris2009}, while previous studies on the fidelity susceptibility only focuses on the intuition behind the geometric interpretation, thus missing the connection with order parameters. The latter is intuitive for symmetry breaking phase transitions where local order parameters are known, and so signatures of the phase transition can be found in reduced states. The most interesting application is in phase transitions without known order parameters, like in our case. This reverses the usefulness argument for the fidelity susceptibility in topological phase transitions: while size scaling of the Fisher information were used to detect transitions without local order parameters, in our case we infer the local/non-local nature of order parameters from the size scaling of the Fisher information.

Basing on the above arguments, we endorse our approach as a powerful tool to characterise general non-equilibrium quantum phase transitions in other systems far beyond previously considered cases, according to the following recipe: superextensivity of the Fisher information, in systems with extensive interactions which scale lineraly with the volume, detects general critical behaviours with at least non-local order parameters, superextensivity of the reduced state Fisher information further proves local order parameters. Our study opens a new avenue of research on NESS-QPT, illustrating that complex structures and relevant features can be extracted by the Fisher information in highly non-trivial systems.

Fisher information is also intimately connected to metrology, being the inverse of the smallest variance in the estimation of the varying parameters \cite{Helstrom1976,Holevo1982,Paris2009}. Superextensivity implies extraordinary enhanced metrological performances. Thus, beyond the aim of NESS-QPT, our study deepens the connection between quantum noisy dynamics and metrology \cite{Bellomo2009,Bellomo2010a,Bellomo2010b,Monras2011,Braun2011,Alipour2014,
Marzolino2014,Marzolino2016-2}, as well as general relations between NESS and quantum information \cite{Albert2016,Marzolino2016}.

The paper is organised as follows. We define the spin chain model with boundary Markovian dissipation in section \ref{model}, and the Fisher information with properties relevant for our analysis in section \ref{Fisher.info}. In section \ref{pert}, we discuss the size scaling of Fisher information for perturbatively small dissipation strength, and implications on non-equilibrium phase transition, including the existence of a critical phase and of (non-)local order parameters. In section \ref{nonpert}, we report on the Fisher information and properties of the non-equilibrium phase transition non-perturbatively in the dissipation strength, and in section   
\ref{conclusions} we conclude.

\section{Spin model} \label{model}

We discuss a $n$-spin chain with XXZ Hamiltonian

\begin{equation}
H_\textnormal{XXZ}=\sum_{j=1}^{n-1}\left(\sigma_j^x\sigma_{j+1}^x+\sigma_j^y\sigma_{j+1}^y+\Delta\sigma_j^z\sigma_{j+1}^z\right),
\end{equation}
which is an archetypical nearest neighbour interaction in condensed matter \cite{Schollwock2004,Franchini2017}, with $\sigma_j^\alpha$ being Pauli matrices of the $j$-th spin. In addition, we consider a uniform magnetic field along the $z$ direction and Markovian dissipation at the boundary of the chain, arriving thus at the following dynamical equation for the density matrix, called master equation,

\begin{eqnarray}
\frac{d}{dt}\rho(t) & = & -i\left[\frac{\Omega}{2}M_z+J H_\textnormal{XXZ},\rho(t)\right] \nonumber \\
& & +\lambda\sum_{k=1}^4\left(L_k\rho(t)L_k^\dag-\frac{1}{2}\left\{L_k^\dag L_k,\rho(t)\right\}\right),
\label{master.eq}
\end{eqnarray}
where

\begin{equation}
L_{1,2}=\sqrt{\frac{1\pm\mu}{2}}\sigma_1^{\pm}, \qquad L_{3,4}=\sqrt{\frac{1\mp\mu}{2}}\sigma_n^{\pm}
\end{equation}
are the so-called Lindblad operators, and $M_z=\sum_{j=1}^n\sigma_j^z$ is the total magnetization along the $z$ direction \footnote{The local Hamiltonian generator $-i[M_z,\cdot]$ commutes with the other terms in \eqref{master.eq}, namely $-i[H_\textnormal{XXZ},\cdot]$ and $L_k\cdot L_k^\dag-\frac{1}{2}\left\{L_k^\dag L_k,\cdot\right\}$. Therefore, the unique NESS $\displaystyle \rho_\infty\equiv\lim_{t\to\infty}\rho(t)$ does not depend on the presence of $-i[M_z,\cdot]$, and equals the NESS derived in the absence of such generator.}.

While the first line of \eqref{master.eq} reproduces the standard Schr\"odinger equation, the second line is the prototypical form of quantum Markovian dissipation, under the minimal physical assumption that the resulting time-evolution $\gamma_t$ be a semigroup, i.e. $\gamma_t\gamma_s=\gamma_{s+t}$ $\forall \, t,s\geqslant 0$, trace preserving, and completely positive, i.e. that preserves positivity of any initial density matrix even when arbitrarily correlated with ancillary systems.

Markovian master equations can be derived from microscopic models with system-environment interaction that is linear in the Lindblad operators
\cite{BreuerPetruccione2002,BenattiFloreanini2005}. In particular, Markovian master equations with local Lindblad operators, i.e. each environment interacting with a single particle as in \eqref{master.eq}, derive from the so-called singular coupling approximation \cite{BreuerPetruccione2002,BenattiFloreanini2005} or from the weak system-environment coupling if the system Hamiltonian is dominated by the interaction-free part \cite{Wichterich2007}, in our case $\Omega\gg J$.

Our model has an exactly solvable steady state density operator, i.e., a fixed point $\rho_\infty=\lim_{t\to\infty}\rho_t$ of \eqref{master.eq}, 
which can be represented in terms of a matrix product ansatz (see Ref.~\cite{Prosen2015} for a review). This structure will be essential to make our computations efficient.

\section{Fisher information} \label{Fisher.info}

Given the aforementioned analytic solution, we compute the Fisher information for variations of the anisotropy $\Delta$

\begin{align}
F_\Delta & =8\lim_{\delta\to0}\frac{1-\sqrt{\mathcal{F}\big(\rho_\infty(\Delta),\rho_\infty(\Delta+\delta)\big)}}{\delta^2}= \nonumber \\
& =2\int_0^\infty ds\,\textnormal{Tr}\left[\left(\frac{\partial\rho_\infty}{\partial\Delta}e^{-s\rho_{\infty}}\right)^2\right], \label{Fisher-info}
\end{align}
with the Uhlmann fidelity $\mathcal{F}(\rho,\sigma)=\big(\textnormal{Tr}\sqrt{\sqrt{\sigma}\rho\sqrt{\sigma}}\big)^2$ \cite{BengtssonZyczkowski2006}.
Defining the eigenvalues $\{p_j\}_j$ and the corresponding eigenvectors $\{|j\rangle\}_j$ of the state $\rho_\infty$, the definition \eqref{Fisher-info} of the Fisher information reads \cite{Helstrom1976,Holevo1982,Paris2009}

\begin{equation} \label{Fisher-info-basis}
F_\Delta=2\sum_{j,l}\frac{\left|\langle j|\partial_\Delta\rho_\infty|l\rangle\right|^2}{p_j+p_l}.
\end{equation}

The connection between Fisher information $F_\Delta$ and estimation theory is summarised in the Cram\'er-Rao bound which bounds any estimation variance of $\Delta$
\cite{Helstrom1976,Holevo1982,Paris2009}. If $\Delta$ is estimated by the measurement of the observable $O$, the Cram\'er-Rao bound reads

\begin{equation} \label{CRB0}
\textnormal{Var}(\Delta)=\frac{\Delta^2 O}{
\left(\frac{\partial}{\partial\Delta}\langle O\rangle\right)^2}\geqslant\frac{1}{F_\Delta},
\end{equation}
where $\Delta^2 O$ is the variance of the observable $O$,
and $\textnormal{Var}(\Delta)$ follows from error propagation.

A property of the Uhlmann fidelity, useful in the following, is that it is non-decreasing under the action of trace preserving and completely positive maps on both the arguments \cite{NielsenChuang}. The partial trace, namely the average over the degrees of freedom of subsystems, is a trace preserving and completely positive map. Therefore, the Fisher information computed from \eqref{Fisher-info} but using reduced states, i.e. resulting from partial traces of the full state $\rho_\infty$, is a lower bound to the Fisher information of $\rho_\infty$. In next sections, we shall use the relation between local order parameters and the Fisher information computed with reduced states instead of full states, that we are going to explain here.

Good order parameters for phase transitions are non-analytic quantities at critical points. Consider local expectations $\langle O\rangle$ with

\begin{equation} \label{loc.OP}
O=\sum_{\mathcal{R}}O_\mathcal{R}, \qquad \langle O_\mathcal{R}\rangle=\textnormal{Tr}\big(\rho_\infty^\mathcal{R}O_\mathcal{R}\big),
\end{equation}
where $\mathcal{R}$ are subsystems with finite, $n$-independent, size,

\begin{equation}
\rho_\infty^{\mathcal{R}}=\textnormal{Tr}_{\bar{\mathcal{R}}}\rho_\infty
\end{equation}
the reduced state resulting from the partial trace over the complement $\bar{\mathcal{R}}$ of
the subsystem $\mathcal{R}$, and $O_\mathcal{R}$ an observable of the subsystem $\mathcal{R}$. Divergences of the derivatives of $\langle O\rangle$ are related to the Fisher information $F_\Delta^\mathcal{R}$ computed
from equation \eqref{Fisher-info} using the state $\rho_\infty^\mathcal{R}$. Suppose that the anisotropy $\Delta$ has to be estimated via measuments of local expectations
$\langle O_\mathcal{R}\rangle$. The Cram\'er-Rao bound is a bound for any estimation variance
\cite{Helstrom1976,Holevo1982,Paris2009}:

\begin{equation} \label{CRB1}
\textnormal{Var}(\Delta)=\frac{\Delta^2 O_\mathcal{R}}{
\left(\frac{\partial}{\partial\Delta}\langle O_\mathcal{R}\rangle\right)^2}\geqslant\frac{1}{F_\Delta^\mathcal{R}},
\end{equation}
where $\Delta^2 O_\mathcal{R}$ is the variance of the observable $O_\mathcal{R}$,
and $\textnormal{Var}(\Delta)$ follows from error propagation. Suppose, instead, to estimate $\Delta$ via experimental measurements of the $k$-th derivative $\frac{\partial^k}{\partial\Delta^k}\langle O_\mathcal{R}\rangle$.
The Cram\'er-Rao bound reads

\begin{equation} \label{CRB2}
\textnormal{Var}(\Delta)=\frac{\textnormal{Var}\left(\frac{\partial^{k-1}}{\partial\Delta^{k-1}}\langle O_\mathcal{R}\rangle\right)}{\left(\frac{\partial^k}{\partial\Delta^k}\langle O_\mathcal{R}\rangle\right)^2}\geqslant\frac{1}{F_\Delta^\mathcal{R}},
\end{equation}
where $\textnormal{Var}\left(\frac{\partial^k}{\partial\Delta^k}\langle O_\mathcal{R}\rangle\right)$ is the variance of the experimental measurements of $\frac{\partial^k}{\partial\Delta^k}\langle O_\mathcal{R}\rangle$.
Such a quantity depends on the measured observables and on instrumental parameters, e.g., if derivatives are estimated via difference quotients, the increment of $\Delta$.
Therefore, the size scaling of the reduced Fisher information bounds from above the degree of divergence of the derivatives of local expectations \eqref{loc.OP}:

\begin{align}
\label{bound.der.loc.OP1}
\left|\frac{\partial}{\partial\Delta}\langle O\rangle\right|\leqslant & \sum_{\mathcal{R}}\left|\frac{\partial}{\partial\Delta}\langle O_{\mathcal{R}}\rangle\right|
\leqslant\sum_{\mathcal{R}}\sqrt{F_\Delta^\mathcal{R} \,
\Delta^2 O_\mathcal{R}
}, \\
\left|\frac{\partial^k}{\partial\Delta^k}\langle O\rangle\right|\leqslant & \sum_{\mathcal{R}}\left|\frac{\partial^k}{\partial\Delta^k}\langle O_{\mathcal{R}}\rangle\right| \nonumber \\
\label{bound.der.loc.OP2}
\leqslant & \sum_{\mathcal{R}}\sqrt{F_\Delta^\mathcal{R} \, \textnormal{Var}\left(\frac{\partial^{k-1}}{\partial\Delta^{k-1}}\langle O_\mathcal{R}\rangle\right)}.
\end{align}
We shall use these bounds to infer the existence of local order parameters.

\section{Pertrubative analysis in the dissipation strength} \label{pert}

We start our investigation with the perturbative analysis for small noise strength $\frac{\lambda}{J}$.
It is worth to stress that this analysis does not correspond to a perturbation around equilibrium. The zeroth order of the NESS is the completely mixed state which does not depend on any parameter. Therefore, there is neither a notion of temperature nor of other equilibrium properties, nor traces of phase transitions in the zeroth order. As a consequence, our perturbative analysis already captures genuine non-equilibrium phase transitions.
In this case, the NESS \cite{Prosen2011a,Prosen2015} is:

\begin{align} \label{NESS.pert}
\rho_\infty= & \frac{1}{2^n}\bigg(\mathbbm{1}+ i\frac{\lambda\mu}{2J}\left(Z-Z^\dag\right) \nonumber \\
& +\frac{\lambda^2\mu}{8J^2}\left(\left[Z,Z^\dag\right]-\mu\left(Z-Z^\dag\right)^2-2\textnormal{Tr}\left(ZZ^\dag\right)\mathbbm{1}\right)\bigg) \nonumber \\
& +\mathcal{O}\left(\frac{\lambda}{J}\right)^3,
\end{align}
where $Z$ is a matrix product operator

\begin{equation} \label{MPO}
Z=\sum_{\{s_1,\dots,s_N\}\in\{0,+,-\}^N}\langle L|\prod_{j=1}^nA_{s_j}|R\rangle\bigotimes_{j=1}^n\sigma_j^{s_j}.
\end{equation}
$A_{s_j}$ are tridiagonal matrices on the auxiliary Hilbert space spanned by the orthonormal basis $\{|L\rangle,|R\rangle,|1\rangle,|2\rangle,\dots,|\lfloor\frac{n}{2}\rfloor\rangle\}$:

\begin{eqnarray}
A_0 & = & |L\rangle\langle L|+|R\rangle\langle R|+\sum_{k=1}^{\lfloor \frac{n}{2}\rfloor}\cos(\eta k)|k\rangle\langle k|, \nonumber \\
A_+ & = & |1\rangle\langle R|-\sum_{k=1}^{\lfloor \frac{n}{2}\rfloor}\sin(\eta k)|k+1\rangle\langle k|, \nonumber  \\
A_- & = & |L\rangle\langle 1|+\sum_{k=1}^{\lfloor \frac{n}{2}\rfloor}\sin(\eta(k+1))|k\rangle\langle k+1|, \label{Apert}
\end{eqnarray}
and $\eta=\arccos\Delta\in\mathbbm{R}\cup i\mathbbm{R}$. The expansion \eqref{NESS.pert} holds as soon as the zeroth order is larger than the first order in $\frac{\lambda}{J}$. Estimating the magnitude of each order with its Hilbert-Schmidt norm ($||O||_{\textnormal{HS}}=\sqrt{\textnormal{Tr}(OO^\dag)}$), the validity condition for \eqref{NESS.pert} reads

\begin{equation} \label{val}
\frac{\lambda}{J}<\frac{\sqrt{2^{n+1}}}{\mu}||Z||_\textnormal{HS}^{-1}=
\begin{cases}
\mathcal{O}\left(\frac{1}{\sqrt{n}}\right) & \textnormal{if } |\Delta|<1, \\
\mathcal{O}\left(\frac{1}{n}\right) & \textnormal{if } |\Delta|=1.
\end{cases}
.
\end{equation}
For $\Delta>1$ the upper bound in \eqref{val} is the inverse of a superexponential function, thus the perturbative expansion \eqref{NESS.pert} is not useful.

\subsection{Non-commuting limits for the Fisher information}

At the lowest order in $\frac{\lambda}{J}$, the Fisher information is

\begin{align} \label{Fisher.info.pert}
F_\Delta & =\frac{\lambda^2\mu^2}{2^{n+1}J^2}\textnormal{Tr}\left(\partial_\Delta Z \, \partial_\Delta Z^\dag\right)+\mathcal{O}\left(\frac{\lambda}{J}\right)^4 \nonumber \\
& =\frac{\lambda^2\mu^2}{J^2}\big(\widetilde F_\Delta+\widehat F_\Delta\big)+\mathcal{O}\left(\frac{\lambda}{J}\right)^4
\end{align}
with the two non-negative contributions

\begin{align}
\label{contributionF1}
\widetilde F_\Delta & =\frac{1}{2(1-\Delta^2)}\sum_{j=1}^n\langle L|\mathbb{A}_0^{j-1}\mathbb{D}\mathbb{A}_0^{n-j}|R\rangle, \\
\label{contributionF2}
\widehat F_\Delta & =\frac{1}{8(1-\Delta^2)}\cdot\frac{d^2}{d\eta^2}\langle L|\mathbb{A}_0^n|R\rangle,
\end{align}
and the matrices $\mathbb{A}_0$ and $\mathbb{D}$ on the auxiliary space of the matrix product structure

\begin{align}
\mathbb{A}_0 = & \!\!\! \sum_{\substack{k,k'=L,R,\\1,\dots,\left\lfloor\frac{n}{2}\right\rfloor}} \!\!\! \left((A_0)_{k,k'}^2+\frac{1}{2}(A_+)_{k,k'}^2+\frac{1}{2}(A_-)_{k,k'}^2\right)|k\rangle\langle k'|, \label{transfer} \\
\mathbb{D} = & \sum_{k=1}^{\lfloor \frac{n}{2}\rfloor}\Big(\textnormal{sign}(1-\Delta^2)\frac{k^2}{2}|k\rangle\langle k|+\frac{k^2}{4}|k+1\rangle\langle k| \nonumber \\
& +\frac{(k+1)^2}{4}|k\rangle\langle k+1|\Big). \label{vertex}
\end{align}
The trace in \eqref{Fisher.info.pert} equals a transition amplitude in the doubled auxiliary space, spanned by $\{|k\rangle\otimes|k'\rangle\}_{k,k'=L,R,1,2,\dots,\lfloor\frac{n}{2}\rfloor}$,
e.g. the left(right)-most state in right-hand-sides of equations \eqref{contributionF1} and \eqref{contributionF2} is actually $\langle L|\otimes\langle L|$ ($|R\rangle\otimes|R\rangle$).
Nevertheless, only the subspace spanned by $\{|k\rangle\otimes|k\rangle\}_{k=L,R,1,2,\dots,\lfloor\frac{n}{2}\rfloor}$ contributes, and then we have applied the mapping $|k\rangle\otimes|k\rangle\to|k\rangle$ to reduce the dimension of the auxiliary space.

Now, we briefly reread results of Refs.~\cite{Marzolino2014,Marzolino2016-2}, originally focused on metrology but not on NESS-QPT, and then we shall report original results in order to end up with a full description of the NESS-QPT.
The system undergoes a NESS-QPT at $|\Delta|=1$, detected by superextensive Fisher information in the leading order

\begin{equation} \label{Fisher.info.pert.Delta1}
F_\Delta\simeq\frac{\lambda^2\mu^2}{32J^2} \, n^4, \qquad \textnormal{for } \frac{\lambda\mu}{J}<\frac{1}{n} \qquad \textnormal{and large } n.
\end{equation}
When the rescaled anisotropy parameter $\frac{\eta}{\pi}=\frac{\arccos\Delta}{\pi}$ is rational and $|\Delta|<1$, the Fisher information in the leading order is

\begin{equation}
F_\Delta\simeq\frac{\lambda^2\mu^2}{J^2}\big(\widetilde\xi \, n^2+\xi \, n\big), \quad \textnormal{for } \frac{\lambda\mu}{J}<\frac{1}{\sqrt{n}} \quad \textnormal{and large } n,
\end{equation}
with size-independent coefficients $\widetilde\xi$ and $\xi$. Thus, $F_\Delta$ cannot be superextensive.
Keeping only the leading contribution of the Fisher information in the thermodynamic limit, and only afterwards setting $\frac{\eta}{\pi}$ to an irrational number results in $F_\Delta=\frac{\lambda^2\mu^2}{J^2}\mathcal{O}(n^5)$, with some oscillations in $n$ damped for more irrational $\frac{\eta}{\pi}$. The latter approach catches the superextensive size scaling of the Fisher information, i.e. the divergent degree of the Fisher information density, when the limit of $\frac{\eta}{\pi}$ approaching irrationals is taken after the thermodynamic limit.

\begin{figure}[htbp]
\centering
\includegraphics[width=\columnwidth]{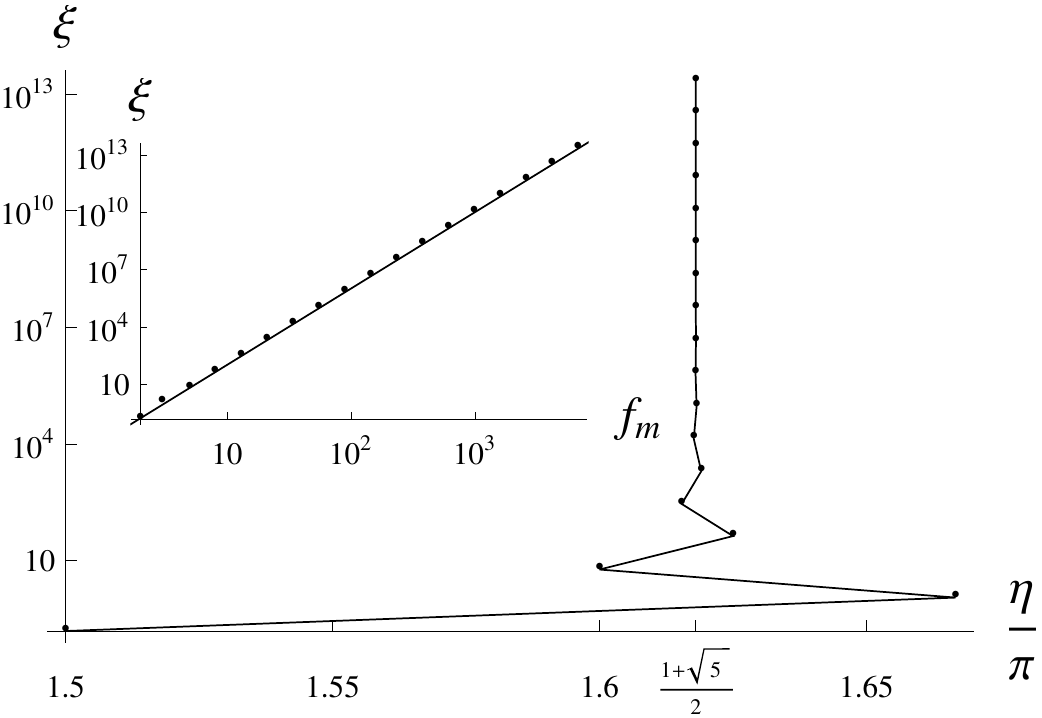}
\caption{Semi-log plot of the coefficient $\xi$ for $\frac{\eta}{\pi}=\frac{f_{m+1}}{f_m}$ with $\{f_m\}_m$ the Fibonacci sequence. The inset shows the log-log plot of $\xi$ as a function of $f_m$, which is perfectly fitted by $(0.0107\pm0.0004)f_m^{3.993\pm0.007}$.}
\label{Fisher-Fibonacci}
\end{figure}

Keeping in mind the just mentioned results of \cite{Marzolino2014,Marzolino2016-2}, we now present original results aiming to complete the characterisation of the NESS-QPT.
We shall show that the limit of $\frac{\eta}{\pi}$ approaching an irrational number does not commute with the thermodynamic limit $n\to\infty$ for $|\Delta|<1$.
Consider first the thermodynamic limit and then the limit of $\frac{\eta}{\pi}$ approaching an irrational number via a sequence of rational approximants, say $\frac{\eta_m}{\pi}=\frac{f_{m+1}}{f_m}$ with $\{f_m\}_m$ the Fibonacci sequence for $m\geqslant3$ which approaches the golden ratio $\varphi=\frac{1+\sqrt{5}}{2}$ as $m\to\infty$. The coefficient $\xi$, plotted in figure \ref{Fisher-Fibonacci}, shows the divergence for $m\to\infty$ fitted by

\begin{equation}
\xi=(0.0107\pm0.0004)f_m^{3.993\pm0.007}
\end{equation}
When $|f_m|\geqslant\left\lfloor\frac{n}{2}\right\rfloor+1$, $\xi(|f_m|)=\xi(\left\lfloor\frac{n}{2}\right\rfloor+1)$,
thus $\xi=\mathcal{O}(n^4)$ in the limit $m\to\infty$ in agreement with the above results.

\begin{figure}[htbp]
\centering
\includegraphics[width=\columnwidth]{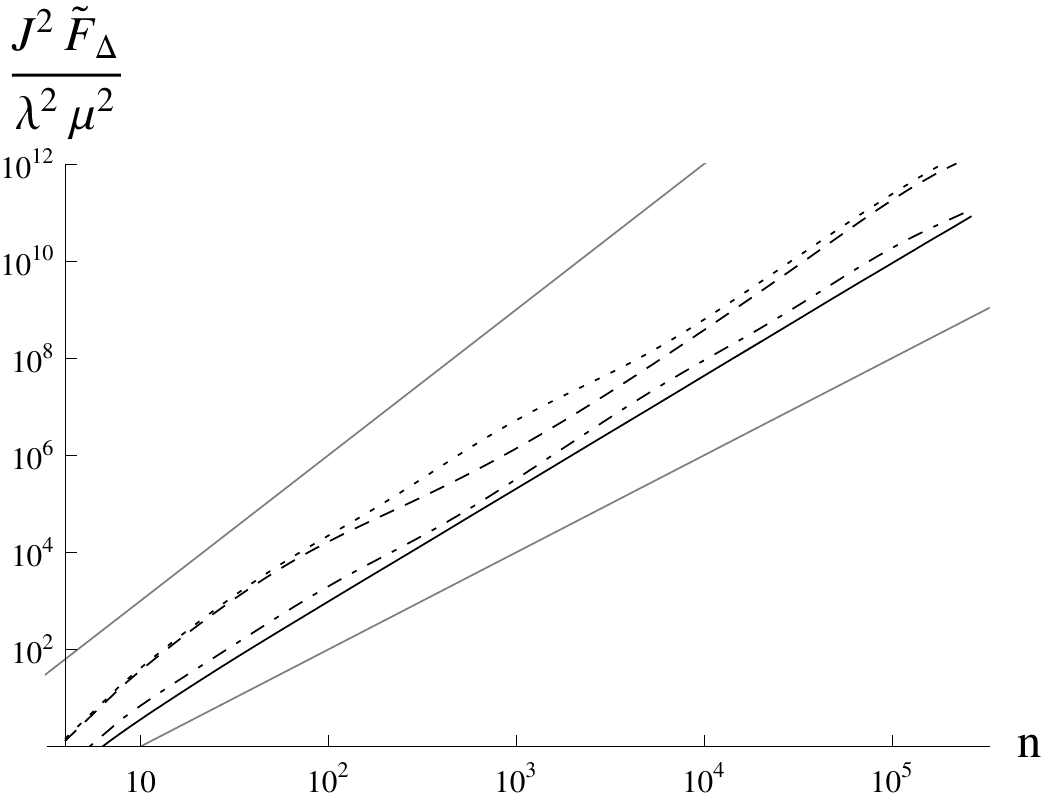}
\caption{Log-log plots of the contribution $\frac{J^2}{\lambda^2\mu^2}\widetilde F_\Delta$ to the Fisher information as function of $n$ for irrational $\frac{\eta}{\pi}$: $\eta=\pi\varphi$ with $\varphi=\frac{1+\sqrt{5}}{2}$ the golden ratio (dark, continuous), $\eta=\pi\frac{\sqrt{3}}{2}$ (dotted), $\eta=\pi^2$ (dashed), $\eta=\pi e$ (dotdashed). For a comparison we also plotted the slopes of power laws $10^{-2}n^2$ and $n^3$ (grey continuous lines).}
\label{Fisher-irrat}
\end{figure}

We now show the numerical computation of the Fisher information with the opposite order of limits, namely at irrational $\frac{\eta}{\pi}$ without any assumption on the particle number $n$.
The log-log plot of the rescaled contribution $\frac{J^2}{\lambda^2\mu^2}\widetilde F_\Delta$ to the Fisher information, with $\widetilde F_\Delta<F_\Delta$, is shown in figure \ref{Fisher-irrat}. We are particularly interested in superextensivity of $F_\Delta$, as a signature of a critical phase, and the remaining contribution to $F_\Delta$, i.e. $\widehat F_\Delta$, can only scale linearly with $n$.
This plot shows
a slower overall growth, as compared to the Fisher information with the limit order exchanged,
with fits given in table \ref{fits.pert}.

\begin{table} [htbp]
\begin{tabular}{|l||l|}
\hline
$\bm{\frac{\eta}{\pi}=}$ & \textbf{fit:} $\bm{\frac{J^2}{\lambda^2\mu^2}\widetilde F_\Delta\simeq}$ \\
\hline
\hline
$\varphi=\frac{1+\sqrt{5}}{2}$ & $(2.112\pm0.002)10^{-2} \, n^{2.32788\pm0.00009}$ \\
\hline
$\frac{\sqrt{3}}{2}$ & $(3.0\pm0.3)10^{-1} \, n^{2.37\pm0.01}$ \\
\hline
$\pi$ & $(1.9\pm0.2)10^{-1} \, n^{2.35\pm0.02}$ \\
\hline
$e$ & $(3.5\pm0.2)10^{-2} \, n^{2.341\pm0.006}$ \\
\hline
\end{tabular}
\caption{Fits of the size scalings plotted in figure \ref{Fisher-irrat}. The fits are more precise when the oscillations are smaller.}
\label{fits.pert}
\end{table}

We have thus shown the non-commutativity of the two limits, but also the superextensivity of the Fisher information for both limit orders, for irrational $\frac{\eta}{\pi}$ which are all critical points.
This indicates that the model has a critical phase for $|\Delta|\leqslant1$ with a highly singular behaviour.

\subsection{Reduced Fisher information and the absence of local order parameters}
\label{reduced.pert}

A critical phase detected by the Fisher information was also observed in the XY model with boundary noise which is mapped to a free Fermion model \cite{Banchi2014}. Our model has Fermion interactions, i.e. the anisotropy term, and the above novel singular behaviour. A critical phase with several peaks of the Fisher information was also found in the topological phase transion of the Kitaev honeycomb model \cite{Yang2008} without local order parameter. This analogy demands a deeper understanding of the NESS-QPT in terms of order parameters. We undertake this investigation basing on the Fisher information of reduced states.

Defining the set $\mathcal{R}=\{\mathcal{R}_j\}_{j=1,\dots|\mathcal{R}|}$ made of $|\mathcal{R}|$ spins at increasing positions $\mathcal{R}_j$, the reduced NESS of this chain portion is

\begin{equation} \label{reduced.NESS.pert}
\rho_\infty^{\mathcal{R}}=\frac{1}{2^{|\mathcal{R}|}}\left(\mathbbm{1}_{2^{|\mathcal{R}|}}+
i\frac{\lambda\mu}{2J}\left(Z_{\mathcal{R}}-Z_{\mathcal{R}}^\dag\right)\right)+\mathcal{O}\left(\frac{\lambda}{J}\right)^2,
\end{equation}
where $\mathbbm{1}_{2^\mathcal{R}}$ is the $2^\mathcal{R}\times 2^\mathcal{R}$ identity matrix, and

\begin{equation} \label{Z_R}
Z_{\mathcal{R}}=
\!\!\!\!\!\!\!
\sum_{\substack{\{s_{\mathcal{R}_j}\}_{j=1,\dots|\mathcal{R}|} \\ \in\{0,+,-\}^{|\mathcal{R}|}}}
\!\!\!\!\!\!\!
\langle L|A_{s_{\mathcal{R}_1}}\prod_{j=2}^{|\mathcal{R}|}A_0^{\mathcal{R}_j-\mathcal{R}_{j-1}-1}A_{s_{\mathcal{R}_j}}|R\rangle
\bigotimes_{j=1}^{|\mathcal{R}|}\sigma_{\mathcal{R}_j}^{s_{\mathcal{R}_j}}.
\end{equation}
In equation \eqref{Z_R}, we have used the fact that $A_0|R\rangle=|R\rangle$ and $\langle L|A_0=\langle L|$.

We shall show upper bounds for the reduced Fisher information and non-increasing $n$ dependence of local expections,
thus of $\Delta^2 O_\mathcal{R}$ and $\textnormal{Var}\left(\frac{\partial^k}{\partial\Delta^k}\langle O_\mathcal{R}\rangle\right)$.
These results, together with equations \eqref{bound.der.loc.OP1} and \eqref{bound.der.loc.OP2}, imply that there are no local order parameters, i.e. non-analytic expectations \eqref{loc.OP}.
For instance, extensive expectations $\langle O\rangle$, e.g. with a number $\mathcal{O}(n)$ of subsystems such as $\mathcal{R}$ labeling single spins or neighbouring couples, cannot have superextensive derivatives.

We start this analysis by bounding the reduced Fisher information of arbitrary subsystems $\mathcal{R}$ with a $n$-independent size $|\mathcal{R}|$ at order $\frac{\lambda^2}{J^2}$:
\begin{equation} \label{bound.reduced.Fisher.info.pert}
F_\Delta^\mathcal{R}\leqslant\mathcal{O}\left(\frac{1}{n}\right)
\end{equation}
which follows from the matrix operator structure of $\rho_\infty$ and $\rho_\infty^\mathcal{R}$ through the following logical steps.

\begin{itemize}
\item Coefficients of $\rho_\infty$ expanded in the tensor basis made of Pauli matrices are generated by products of
sequences of $n$ tridiagonal matrices on an auxiliary space \cite{Marzolino2016}, as shown in equation \eqref{MPO}.
\item The dependence of these coefficients on $n$ enters through the number of matrices in the sequence generating a $\lfloor\frac{n}{2}\rfloor$-dimensional
auxiliary subspace.
\item The coefficients of $\rho_\infty^\mathcal{R}$ in the Pauli tensor basis have analogous structure, as shown in equation \eqref{Z_R},
but with the diagonal matrix $A_0$ in the matrix product at positions corresponding to traced-out spins.
\item This diagonal matrix $A_0$ does not have raising and lowering operators, and thus the dimension of the generated auxiliary subspace equals $\frac{|\mathcal{R}|}{2}$.
As a consequence, the dependence of $\rho_\infty^\mathcal{R}$ on $n$ is manifest only from the exponents $\mathcal{R}_j-\mathcal{R}_{j-1}+1<n$.
\item The modulus of $A_0$ is strictly upper bounded by the identity matrix at $|\Delta|<1$.
Therefore, the coefficients of $\rho_\infty^\mathcal{R}$ in the Pauli tensor basis are upper bounded in modulus by an exponentially decaying function due to $A_0^{\mathcal{R}_j-\mathcal{R}_{j-1}-1}$
if some $\mathcal{R}_j-\mathcal{R}_{j-1}-1$ grow with $n$, or by an $n$-independent contribution if $\mathcal{R}_j-\mathcal{R}_{j-1}-1=\mathcal{O}(n^0)$ for $j\in\big[2,|\mathcal{R}|\big]$.
This already proves that local expections, and thus $\Delta^2 O_\mathcal{R}$ and $\textnormal{Var}\left(\frac{\partial^k}{\partial\Delta^k}\langle O_\mathcal{R}\rangle\right)$, do not increase with $n$.
The above $n$-dependence, together with the definition \eqref{Fisher-info} and the range $\frac{\lambda}{J}<\frac{1}{\sqrt{n}}$ of the perturbation expansion, implies the bound \eqref{bound.reduced.Fisher.info.pert} for $|\Delta|<1$.
\item At $|\Delta|=1$, the eigenvalues of $A_0$ are $\pm1$, and local expections remain non-increasing with $n$. Furthermore, the derivative in the definition \eqref{Fisher-info} gives an additional multiplicative factor upper bounded by $n$,
when deriving the term $A_0^{\mathcal{R}_j-\mathcal{R}_{j-1}-1}$, but with the exponential damping suppressed in the limit $|\Delta|\to1$.
This multiplicative factor is however insufficient to compensate the smallness of $\frac{\lambda}{J}<\frac{1}{n}$ which is the validity range of the perturbation expansion at $|\Delta|=1$. This again implies the bound \eqref{bound.reduced.Fisher.info.pert}.
\end{itemize}

Remarkable examples are reduced states $\rho_\infty^\mathcal{R}$ of contiguous blocks of spins, i.e. $\mathcal{R}=[\mathcal{R}_{\textnormal{min}},\mathcal{R}_{\textnormal{max}}]$, so with all traced-out spins near the boundaries. These reduced states, i.e. equation \eqref{reduced.NESS.pert} with $\mathcal{R}_j-\mathcal{R}_{j-1}-1=0$, have exactly the same analytic form of the full steady state $\rho_\infty$ with $n$ replaced by the number of spins $|\mathcal{R}|=\mathcal{R}_{\textnormal{max}}-\mathcal{R}_{\textnormal{min}}$ in the subsystem. The system is thus selfsimilar.

Summing up, $k$-th derivatives of any observable with $k=\mathcal{O}(n^0)$ lack divergent behaviours. If both $k$ and $\mathcal{R}_j-\mathcal{R}_{j-1}-1$ grow with $n$, $k$-th derivatives can diverge because of repeated derivations of terms $A_0^{\mathcal{R}_j-\mathcal{R}_{j-1}-1}$.
Since the lower bound in \eqref{CRB2} does not depend on $k$, this divergence must be compensated by the numerator of the left-hand-side. Intuitively, to measure derivatives at increasing orders,
e.g. via different quotients, we need to distinguish many measurements all at values of $\Delta$ that lie within a very, ideally vanishingly, small interval. Thus, the measurements of such derivatives become very hard, witnessed by large $\textnormal{Var}\left(\frac{\partial^k}{\partial\Delta^k}\langle O_\mathcal{R}\rangle\right)$, and so is the consequent determination of $\Delta$ as imposed by the Cram\'er-Rao bound \eqref{CRB2}.


The above discussion is insufficient for the reduced states of a single spin, i.e. $\mathcal{R}={\{j\}}$, because it is completely mixed at order $\frac{\lambda}{J}$, i.e. $\rho_\infty^{\mathcal{R}=\{j\}}=\frac{\mathbbm{1}_2}{2}+\mathcal{O}\left(\frac{\lambda}{J}\right)^2$, and the next order is

\begin{equation} \label{red.state.pert}
\rho_\infty^{\mathcal{R}=\{j\}}=\frac{1}{2}\left(\mathbbm{1}_2+\frac{\lambda^2\mu}{4J^2}\gamma_j\sigma_j^z\right)+\mathcal{O}\left(\frac{\lambda}{J}\right)^3,
\end{equation}
with

\begin{align} \label{gamma.pert}
\gamma_j= & \langle L|\mathbb{A}_0^{j-1}\mathbb{A}_z\mathbb{A}_0^{n-j}|R\rangle, \\
\mathbb{A}_z= & \frac{1}{2}\sum_{\substack{k,k'=L,R,\\1,\dots,\left\lfloor\frac{n}{2}\right\rfloor}}\left((A_+)_{k,k'}^2-(A_-)_{k,k'}^2\right)|k\rangle\langle k'|,
\end{align}
and $\mathbb{A}_0$ defined in equation \eqref{transfer}. As before, we have applied the mapping $|k\rangle\otimes|k\rangle\to|k\rangle$ to reduce the dimension of the auxiliary space because only the subspace spanned by $\{|k\rangle\otimes|k\rangle\}_{k=L,R,1,2,\dots,\lfloor\frac{n}{2}\rfloor}$ contributes.

The corresponding single spin reduced Fisher information is

\begin{equation}
F_\Delta^{\mathcal{R}=\{j\}}=\frac{\left(\frac{\partial}{\partial\Delta}\langle\sigma_j^z\rangle\right)^2}{\Delta^2\sigma_j^z}=
\frac{\lambda^4\mu^2}{256 J^4}\left(\frac{\partial\gamma_j}{\partial\Delta}\right)^2+\mathcal{O}\left(\frac{\lambda}{J}\right)^5.
\end{equation}
From numerical computations, we found intensive, and even very small $F_\Delta^{\mathcal{R}=\{j\}}$ at $|\Delta|<1$. This, together with the validity condition $\frac{\lambda}{J}<\frac{1}{\sqrt{n}}$ of the perturbative expansion at $|\Delta|<1$, implies a bound tighter than \eqref{bound.reduced.Fisher.info.pert}, namely

\begin{equation} \label{reduced.1spin.Fisher.next}
F_\Delta^{\mathcal{R}=\{j\}}\leqslant\mathcal{O}\left(\frac{\lambda^4}{J^4}n^0\right)<\mathcal{O}\left(\frac{1}{n^2}\right).
\end{equation}

The zeroth and the first orders of $\gamma_j$ around $|\Delta|=1$ can be analytically computed truncating the auxiliary space to the subsystem spanned by $\{|L\rangle,|R\rangle,|1\rangle,|2\rangle\}$:

\begin{equation}
\left.\gamma_j\right|_{\Delta=\pm1}=\frac{1}{4}(n-2j+1)\big(1\pm(n-2)(\Delta\mp 1)\big)+\mathcal{O}(\Delta\mp 1)^2.
\end{equation}
Therefore, the Fisher information of the reduced state \eqref{red.state.pert} at $|\Delta|=1$ is

\begin{equation}
F_{\Delta=\pm1}^{\mathcal{R}=\{j\}}=\frac{\lambda^4\mu^2}{256 J^4}(n-2)^2(n-2j+1)^2+\mathcal{O}\left(\frac{\lambda}{J}\right)^5.
\label{reduced.1spin.Fisher.next.Delta1}
\end{equation}
The Fisher information of the single spin reduced state \eqref{reduced.1spin.Fisher.next.Delta1} exhibits an apparent superextensive behaviour, i.e. a power law size scaling with exponent between $2$ and $4$, depending on the spin position $j$. Nevertheless, this power law is reduced by the validity range of the perturbative expansion, namely $\frac{\lambda}{J}<\frac{1}{n}$ at $|\Delta|=1$. Therefore the bound to the reduced Fisher information is

\begin{equation}
F_\Delta^{\mathcal{R}=\{j\}}\leqslant\mathcal{O}\left(\frac{\lambda^4}{J^4}n^4\right)<\mathcal{O}(n^0).
\end{equation}
Although the above bound does not allow for local order parameters, its increase with respect to \eqref{bound.reduced.Fisher.info.pert} suggests that superextensivity of $F_{\Delta=\pm1}^{\mathcal{R}={\{j\}}}$ might gradually emerge when inceasing the order of $\frac{\lambda}{J}$ and at non-perturbative regime, as we will discuss in section \ref{nonpert}.

These results imply that there are no local order parameters detecting the NESS-QPT, as for tolopogical phase transitions.

\subsection{Non-local order parameters}

Although there are no local order parameters at the lowest order in $\frac{\lambda}{J}$, there are non-local order parameters which detect at least the onset of the critical phase $|\Delta|=1$, for instance the expectation of

\begin{equation} \label{O_Delta}
O_\Delta=\left.2^{n+1}\frac{J}{\mu}\cdot\frac{\partial}{\partial\lambda}\rho_\infty\right|_{\lambda=0}
\end{equation}
or its limit $O_{\Delta\to\pm1}$ if one prefers a $\Delta$-independent operator.
The expectation of \eqref{O_Delta} satisfies

\begin{align}
\langle O_\Delta\rangle& =\frac{\lambda\mu}{J}\langle L|\mathbb{A}_0^n|R\rangle+\mathcal{O}\left(\frac{\lambda}{J}\right)^2 \nonumber \\
& \xrightarrow[\Delta\to\pm1]{}\frac{\lambda\mu}{8J}n(n-1)+\mathcal{O}\left(\frac{\lambda}{J}\right)^2
\end{align}
and

\begin{equation}
\frac{\partial}{\partial\Delta}\langle O_\Delta\rangle\xrightarrow[\Delta\to\pm1]{}\mp\frac{\lambda\mu}{12J}n(n-1)(n-2)+\mathcal{O}\left(\frac{\lambda}{J}\right)^2.
\end{equation}
The variance of $O_\Delta$ is

\begin{align}
\Delta^2O_\Delta & =2\langle L|\mathbb{A}_0^n|R\rangle+\mathcal{O}\left(\frac{\lambda}{J}\right) \nonumber \\
& \xrightarrow[\Delta\to\pm1]{}\frac{1}{4}n(n-1)+\mathcal{O}\left(\frac{\lambda}{J}\right).
\end{align}
The non-local order parameter has then superextensive derivative, i.e. divergent density of the derivative in the thermodynamic limit. The density derivative is $\frac{1}{n}\frac{\partial}{\partial\Delta}O_\Delta=\mathcal{O}(n^{2-\alpha})$ for $\frac{\lambda\mu}{J}=\mathcal{O}\big(\frac{1}{n^\alpha}\big)$ with $\alpha\in(1,2)$, compatibly with the range of validity $\frac{\lambda\mu}{J}<\frac{1}{n}$ of the perturbative expansion for $|\Delta|=1$.
The ratio $\left(\frac{\partial}{\partial\Delta}\langle O_\Delta\rangle\right)^2/\Delta^2O_\Delta$ has also the same scaling of the Fisher information $F_\Delta\big|_{|\Delta|=1}$ \eqref{Fisher.info.pert.Delta1} almost saturating the Cram\'er-Rao bound \eqref{CRB0} with $O=O_\Delta$.

\section{Non-pertrubative analysis in the dissipation strength} \label{nonpert}

In order to investigate the non-perturbative behaviour of the Fisher information, we consider the steady state of the master equation \eqref{master.eq} with $\mu=1$ which is known for any $\lambda$ \cite{Prosen2011b,Karevski2013,Popkov2013,Prosen2015}:

\begin{equation} \label{MPO2}
\rho_\infty=\frac{SS^\dag}{\textnormal{Tr}(SS^\dag)}, \quad S=\sum_{\substack{\{s_1,\dots,s_n\}\\\in\{0,+,-\}^n}}\langle 0|\prod_{j=1}^nB_{s_j}|0\rangle\bigotimes_{j=1}^n\sigma_j^{s_j},
\end{equation}
with the matrix product operator $S$ and tridiagonal matrices $B_{s_j}$ on the auxiliary Hilbert space spanned by the orthonormal basis $\{|0\rangle,|1\rangle,|2\rangle,\dots,|\lfloor\frac{n}{2}\rfloor\rangle\}$

\begin{eqnarray}
B_0 & = & \sum_{k=0}^{\lfloor\frac{n}{2}\rfloor}\frac{\sin(\eta(s-k))}{\sin(\eta s)}|k\rangle\langle k|, \nonumber \\
B_+ & = & -\sum_{k=0}^{\lfloor\frac{n}{2}\rfloor}\frac{\sin(\eta(k+1))}{\sin(\eta s)}|k\rangle\langle k+1|, \nonumber \\
B_- & = & \sum_{k=0}^{\lfloor\frac{n}{2}\rfloor}\frac{\sin(\eta(2s-k))}{\sin(\eta s)}|k+1\rangle\langle k|, \label{Bnonpert}
\end{eqnarray}
and with $s$ given by $8i\sin\eta\cot(s\eta)=\lambda$.

While the numerical or analytical computation of the full state Fisher information \eqref{Fisher-info} is very hard, the computation of reduced states of small subsytems and their reduced Fisher information is feasible.
The reduced Fisher information is a lower bound of the full state Fisher information because the Uhlmann fidelity is a non-decreasing function under the action of trace preserving and completely positive maps, like the partial trace, on both the arguments \cite{NielsenChuang}.
Therefore, superextensivity of the reduced Fisher information immediately implies superextensivity of the full state's Fisher information, which is the more general signature of the phase transition. As explained in section \ref{Fisher.info}, superextensivity of the reduced Fisher information also provides additional knowledge, e.g. proving the existence and deriving local order parameters.

The $j$-th spin reduced state is diagonal in the $\sigma_j^z$ basis:

\begin{align}
\label{1spin.rho.nonpert} \rho_\infty^{(j)} & =\frac{1}{2}\left(\mathbbm{1}+\gamma_j\sigma_j^z\right), \\
\label{gamma.nonpert} \gamma_j & =\langle\sigma_j^z\rangle=\frac{\langle 0|\mathbb{B}_0^{j-1}\mathbb{B}_z\mathbb{B}_0^{n-j}|0\rangle}{\langle 0|\mathbb{B}_0^n|0\rangle},
\end{align}
and

\begin{align}
\mathbb{B}_0= & \sum_{k,k'=0}^{\left\lfloor\frac{n}{2}\right\rfloor}\left(\left|(B_0)_{k,k'}\right|^2+\frac{1}{2}\left|(B_+)_{k,k'}\right|^2+\frac{1}{2}\left|(B_-)_{k,k'}\right|^2\right) \nonumber \\
& \times|k\rangle\langle k'|, \\
\mathbb{B}_z= & \frac{1}{2}\sum_{k,k'=0}^{\left\lfloor\frac{n}{2}\right\rfloor}\left(\left|(B_+)_{k,k'}\right|^2-\left|(B_-)_{k,k'}\right|^2\right)|k\rangle\langle k'|.
\end{align}
We have again applied the mapping $|k\rangle\otimes|k\rangle\to|k\rangle$ to reduce the dimension of the auxiliary space because only the subspace spanned by $\{|k\rangle\otimes|k\rangle\}_{k=0,1,2,\dots,\lfloor\frac{n}{2}\rfloor}$ contributes.

\begin{figure}[htbp]
\centering
\includegraphics[width=\columnwidth]{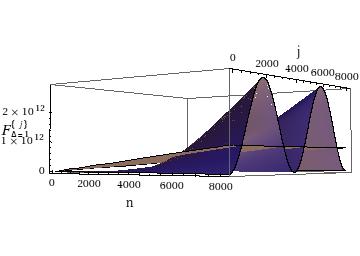}
\includegraphics[width=\columnwidth]{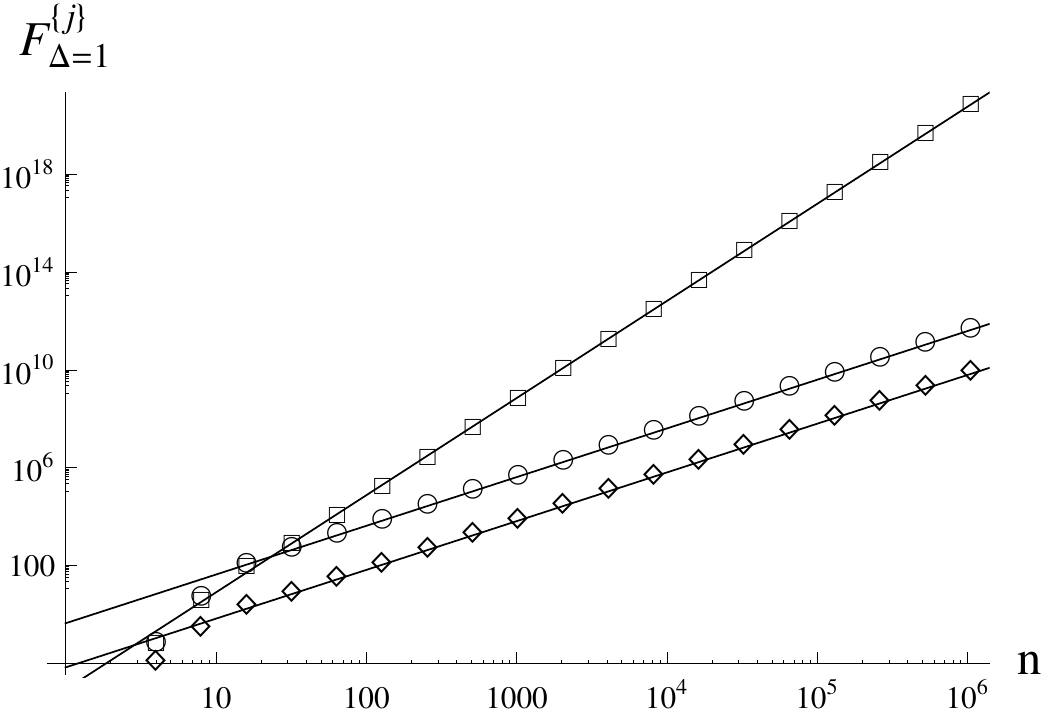}
\caption{Upper panel: Plot of the Fisher information $F_{\Delta=1}^{\{j\}}$ of the $j$-th spin reduced state as a function of $n$ and $j$. The superextensivity is manifest from the comparison with the plane $10^8 n$. Lower panel: Log-log plots of the $F_{\Delta=1}^{\{j\}}$ as function of $n$, set to powers of $2$, for $\lambda=1$ and $j=1$ (circles), $j=\left\lfloor\frac{n}{4}\right\rfloor$ (squares), and $j=\left\lfloor\frac{n}{2}\right\rfloor$ (diamonds). The continuous lines are the corresponding fits, see figure \ref{fits.nonpert}, excluding the first three points of each line, which clearly deviates from the large $n$ behaviour.}
\label{reduced-Fisher}
\end{figure}

Therefore, the $j$-th spin reduced Fisher information is

\begin{equation} \label{red.Fisher.info.nonpert}
F_\Delta^{\mathcal{R}=\{j\}}=\frac{\left(\frac{\partial}{\partial\Delta}\langle\sigma_j^z\rangle\right)^2}{\Delta^2\sigma_j^z},
\end{equation}
saturating the Cram\'er-Rao bound \eqref{CRB1} with $O_{\mathcal{R}={\{j\}}}=\sigma_j^z$.
The derivative $\frac{\partial}{\partial\Delta}\langle\sigma_j^z\rangle$ and $F_\Delta^{\{j\}}$ are both superextensive at $|\Delta|=1$, as shown in figure \ref{reduced-Fisher} for $\Delta=1$.
The superextensive size scalings of $F_\Delta^{\{j\}}$ are fitted with power laws listed in table \ref{fits.nonpert}.
The case $\Delta=-1$ gives similar results.
The reduced Fisher information $F_\Delta^{\{j\}}$ is also symmetric with respect to reflection of the spin chain around its center.

\begin{figure}[htbp]
\centering
\includegraphics[width=\columnwidth]{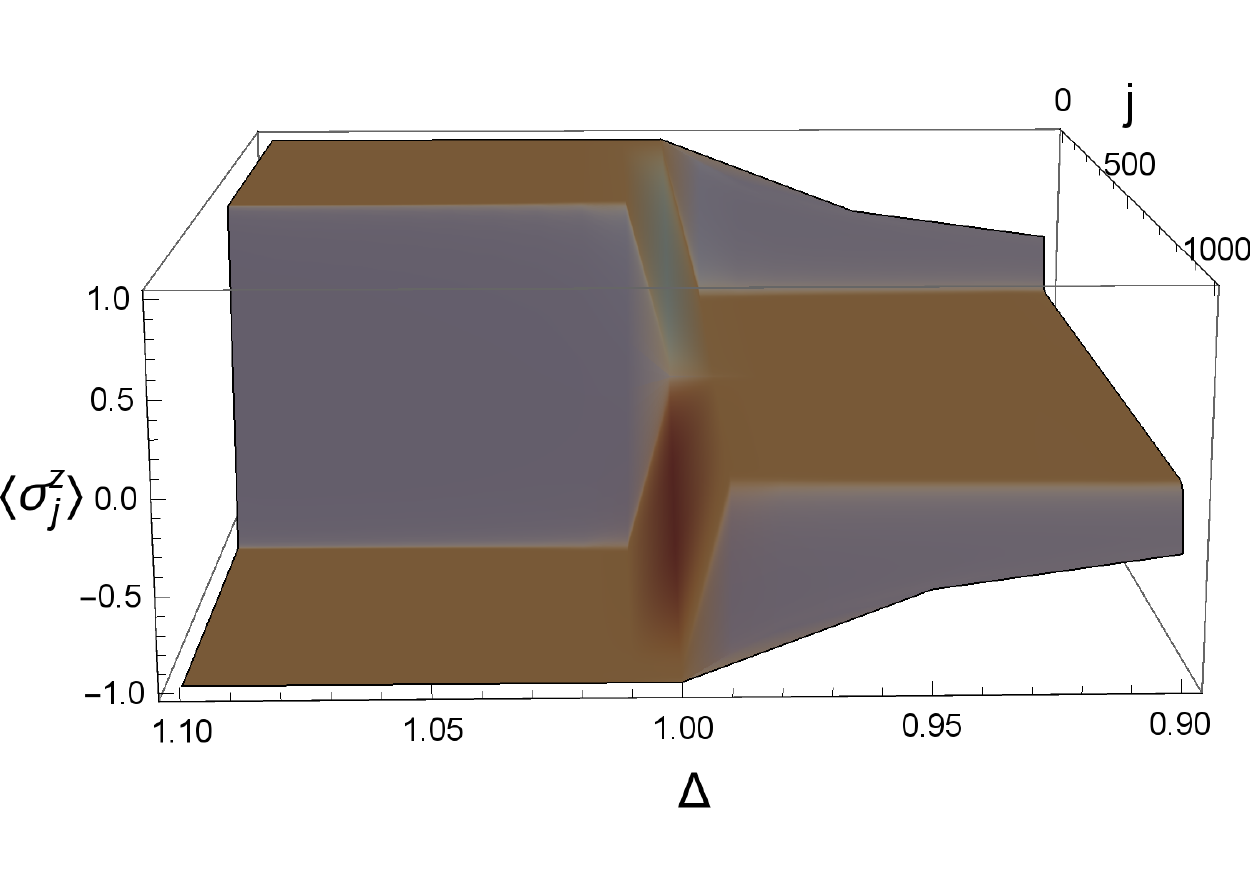}
\caption{Plot of the magnetisation profile $\langle\sigma_j^z\rangle$ of the $j$-th spin as a function of $j$ and $\Delta$ for $n=1000$ and $\lambda=1$.}
\label{magn.profile}
\end{figure}

As a consequence of the above superextensivity, the magnetisation profile $\langle\sigma_j^z\rangle$ is an intensive, local order parameter for the critical points $|\Delta|=1$,
with diverging derivative $\frac{\partial}{\partial\Delta}\langle\sigma_j^z\rangle$, plotted in figure \ref{magn.profile}.
The finite size scaling of $\frac{\partial}{\partial\Delta}\langle\sigma_j^z\rangle$ equals the square root of that of the reduced Fisher information $F_\Delta^{\mathcal{R}=\{j\}}$,
from \eqref{red.Fisher.info.nonpert} because the variance in the denominator is $\Delta^2\sigma_j^z=1-\langle\sigma_j^z\rangle^2=\mathcal{O}(n^0)$ in agreement with the plot in figure \ref{magn.profile}.
Extensive local order parameters are the magnetisations $\sum_{j\in\mathcal{R}}\langle\sigma_j^z\rangle$ for any macroscopic but not centrosymmetric portion $\mathcal{R}$ of the chain.
For centrosymmetric portions, the divergences at spin positions $j$ and $n-j$ cancel with each other.
Other extensive local order parameters are $\sum_{j\in\mathcal{R}}f\left(\langle\sigma_j^z\rangle\right)$ with even functions $f(\cdot)$ and for any set $\mathcal{R}$, even centrosymmetric ones.

\begin{table} [htbp]
\begin{tabular}{|l||l|}
\hline
\textbf{spin position:} $\bm{j=}$ & \textbf{fit:} $\bm{F_{\Delta=1}^{\{j\}}\simeq}$ \\
\hline
\hline
$1$ & $(4.36\pm0.06)10^{-1} \, n^{1.992.\pm0.002}$ \\
\hline
$\left\lfloor\frac{n}{4}\right\rfloor$ & $(8.5\pm0.9)10^{-4} \, n^{3.97.\pm0.01}$ \\
\hline
$\left\lfloor\frac{n}{2}\right\rfloor$ & $(6.8\pm0.2)10^{-3} \, n^{1.993.\pm0.003}$ \\
\hline
\end{tabular}
\caption{Fits of the size scalings plotted in figure \ref{reduced-Fisher}.}
\label{fits.nonpert}
\end{table}

The reduced Fisher information does not show superextensive size scaling at $|\Delta|<1$.
Therefore, the superextensivity of the full state Fisher information at $|\Delta|<1$, and thus the presence of a critical phase in the non-perturbative regime, is still an open question.

\section{Conclusions} \label{conclusions}

We derived characterizations of the NESS-QPT of the XXZ model with boundary noise, starting from the Fisher information. We identified a critical phase defined by the anisotropy range $|\Delta|\leqslant1$, with irrational $\frac{\eta}{\pi}$ being critical points, for small dissipation. For instance, we observed a clear divergence for $\frac{\eta}{\pi}$ approaching the golden ratio through the Fibonacci sequence, and superextensive Fisher information at different irrational $\frac{\eta}{\pi}$. This critical behaviour lacks local order parameters but exhibits non-local ones. Moreover, it is observed for a small dissipation strength which vanishes for infinite particle number. This limit might be considered similar to reducing the XYZ model to the XY model which yet exhibits a phase transition. Moreover, other topological characterisations of phase transitions already revealed critical points with non-analytic microcanonical entropy at finite size which becomes smoother as the particle number grows and analytic in the thermodynamic limit \cite{Pettini2007,Kastner2008}.

At non-perturbative dissipation, the reduced Fisher information provides a superextensive lower bound to the full state Fisher information at $|\Delta|=1$ together with local order parameters, e.g. the magnetisation profile.
Since the reduced Fisher information of the non-perturbative NESS is not superextensive for $|\Delta|<1$, it is still an open question whether the Fisher information is superextensive.

We have proved the power of the Fisher information approach to characterise NESS-QPT. We suggest that this approach will be useful for many other critical phenomena, such as classical non-equilibrium phase transitions \cite{HenkelHinrichsenLubeck2008,HenkelHinrichsenLubeck2010,Odor2004,Lubeck2004}, in quenched and dynamical systems \cite{Heyl2013,Vajna2015}, or in chaotic systems \cite{Pettini2007,Wimberger2016}. Superextensive Fisher information also identifies probes for the estimation of the control parameter with enhanced performances \cite{Helstrom1976,Holevo1982,Paris2009}. Our system, having a NESS with a very low or vanishing entanglement, is also relevant for enhanced metrological schemes without entanglement \cite{BraunReview}.

{\bf Acknowledgments.} The work has been supported by ERC grant OMNES, the grants P1-0044 and J1-5439 of Slovenian Research Agency, and by H2020 CSA Twinning project No.~692194, ``RBI-T-WINNING''. U.M. warmly thanks Marko Ljubotina and Marko \v{Z}nidari\v{c} for their useful advice on numerical computations.

\end{document}